\documentclass[aps,prd,onecolumn,superscriptaddress,showpacs,floatfix,preprintnumbers]{revtex4-2}
\usepackage[dvipsnames]{xcolor}
\usepackage[english]{babel}
\usepackage[T1]{fontenc}
\usepackage{lmodern}
\usepackage{mathtools}
\usepackage{relsize}
\usepackage{amssymb}
\usepackage{slashed}
\usepackage{enumerate}
\usepackage{appendix}
\makeatletter
\g@addto@macro\bfseries{\boldmath}
\makeatother

\definecolor{lcolor}{rgb}{0.5,0,0}
\definecolor{citcolor}{rgb}{0,0,1}
\usepackage[breaklinks,colorlinks,urlcolor=blue,citecolor=citcolor,linkcolor=lcolor,linktoc=all]{hyperref}

\renewcommand{\tilde}{\widetilde}
\renewcommand{\Re}{\operatorname{Re}}
\renewcommand{\Im}{\operatorname{Im}}
\newcommand{\Sec}{Sec.~}

\begin{document}
\title{Augmenting the residue theorem with boundary terms in finite-density calculations}
\preprint{HIP-2022-21/TH}
\author{Tyler Gorda}
\email{tyler.gorda@physik.tu-darmstadt.de}
\affiliation{Technische Universit\"{a}t Darmstadt, Department of Physics, 64289 Darmstadt, Germany}
\affiliation{ExtreMe Matter Institute EMMI and Helmholtz Research Academy for FAIR, GSI Helmholtzzentrum f\"ur Schwerionenforschung GmbH, 64291 Darmstadt, Germany}
\author{Juuso Österman}
\email{juuso.s.osterman@helsinki.fi}
\affiliation{Department of Physics and Helsinki Institute of Physics,
P.O.~Box 64, FI-00014 University of Helsinki, Finland}
\author{Saga Säppi}
\email{saga.saeppi@tum.de}
\affiliation{European Centre for Theoretical Studies in Nuclear Physics and Related Areas (ECT*) and Fondazione Bruno Kessler, Strada delle Tabarelle 286, I-38123, Villazzano (TN), Italy}

\begin{abstract}
\noindent
At zero temperature and finite chemical potential, $d$-dimensional loop integrals with complex-valued integrands in the imaginary-time formalism yield results dependent on the integration order. We observe this even with the simplest one-loop dimensionally regularized integrals. Computing such integrals by evaluating the spatial $\mathrm{d}^{d} p$ integral before the temporal $\mathrm{d} p_0$ integral yields results consistent with those obtained at small but nonvanishing temperatures. Computing the temporal integral first by applying the residue theorem to the integrand yields a different answer. The same holds for general complexified propagators. In this work we aim to understand the theoretical background behind this difference, in order to fully enable the powerful techniques of residue calculus in applications. We cast the difference into the form of a derivative term related to Dirac deltas, and further demonstrate how the difference originates from the  zero-temperature limit of the Fermi--Dirac occupation functions treated as complex-valued functions. We also discuss a generalization to propagators raised to non-integer powers.

\end{abstract}

\maketitle

\section{Introduction}
\noindent
Loop integrals appearing in the perturbative expansion of quantum field theories are often ill-defined when interpreted as ordinary integrals \cite{peskin, Collins:1984xc, Weinberg:1995mt}. In order to evaluate them, a number of different regulators have been introduced, with the most commonly seen modern method being dimensional regulation \cite{hooft, Bollini}. Using this formalism, an integral is formally evaluated in $d+1$ spacetime dimensions with $d = 3 - 2 \epsilon$ taking values in some nonempty open set, and the result is analytically continued to obtain a result near physically relevant values of $d$. 

Such integrals are often manipulated somewhat carelessly, with hope that the analytical continuation justifies freely changing the order of integration, and correctly handles any issues arising from divergences and discontinuities. In many cases, these manipulations still lead to the correct answer, but there are interesting exceptions. One particular class of exceptions occurs within thermal field-theory calculations at zero temperature $T = 0$ and finite chemical potentials $\mu > 0$, as we shall now illustrate. While finite chemical potential will be our main focus as far as applications go, the subtleties discussed here have at least a possibility to arise in any loop integrals involving complex propagators. This can happen for example in studies of decay processes and other calculations dealing with complex-valued momenta (for example amplitude calculations are well-known to use complex kinematics, see \cite{Britto:2005fq}.)

High-density calculations at $T = 0$ are most properly viewed as $T \to 0$ limits of $T > 0$ computations. However, the latter are often much more complicated, as they depend on two dimensionful parameters $T$ and $\mu$, rather than on one. For this reason, many high-density field theory calculations (such as those in cold quark matter \cite{Kurkela:2009gj,Gorda:2021znl,Gorda:2021kme}) are performed at exactly $T = 0$, and often proceed using repeated application of the residue theorem from complex analysis, which allows one to arrive at final expressions after proceeding through significantly simpler intermediate steps. What one ignores in this procedure is that the finite temperature expressions often have the temperature $T$ as a regulator, and as such, switching to a less-regulated $T = 0$ expression may induce problems.

For the purpose of illustration, consider the following simple one-loop integral with no numerator structure, given in the imaginary-time formalism and regulated in the $\overline{\text{MS}}$-scheme
\begin{equation}\label{eq:Iadef}
    I_\alpha(\mu) \equiv\int_{-\infty}^\infty \frac{\text{d} p_0}{2 \pi} \int_p  \frac{1}{\left[(p_0+i\mu)^2 + p^2\right]^\alpha},
\end{equation}
where $\alpha \in \mathbb{R}_+$, and where we have defined the $d$-dimensional spatial integral 
\begin{equation}
    \int_p = \left( \frac{ e^{\gamma_\text{E}} \Lambda^2}{ 4 \pi} \right)^\epsilon \int_{\mathbb{R}^d} \frac{\text{d}^d\mathbf{p}  }{(2\pi)^d} = \left( \frac{ e^{\gamma_\text{E}} \Lambda^2}{ 4 \pi}\right)^\epsilon \frac{2}{(4\pi)^{d/2}\Gamma(d/2)} \int_{\mathbb{R_+}} \text{d}p \,  p^{d-1}.
\end{equation}
Let us emphasize here that even non-integer values of $\alpha$ are of physical interest. In multi-loop computations in finite-density quantum field theory, such integrals can arise for example in cases where a loop integral is written as an iterated integral, as is seen explicitly in e.g.~\cite{Vuorinen:2003fs}, Eqs.~(B.62), (B.63).

An extension of $I_{\alpha}(\mu)$ to finite temperatures $T > 0$ (in which the integral over the $0$ component is replaced by a sum over discrete Matsubara modes) has been evaluated in literature \cite{Vuorinen:2003fs}. Evaluating the corresponding sum-integral in $d=3-2\epsilon$ dimensions (again in the $\overline{\text{MS}}$-scheme) by first computing the spatial integral over $p$ results in
\begin{equation}
\begin{split}
    I_{\alpha}(\mu,T) &\equiv T \sum_{\{\omega_n\} } \int_p \frac{1}{\left[ p^2 + \left( \omega_n+i \mu \right)^2\right]^\alpha} \\
    &=\left(  \frac{e^{\gamma_\text{E}} \Lambda^2}{4 \pi}\right)^\epsilon
    \frac{T (2 \pi T)^{d-2\alpha}}{(4\pi)^{d/2}} 
    \frac{\Gamma\left(\alpha-\frac{d}{2}\right)}{\Gamma\left(\alpha\right)} \Big[ \zeta\left(2\alpha-d,\frac{1}{2}-i\frac{\mu}{2\pi T}\right)+\zeta\left(2\alpha-d,\frac{1}{2}+i\frac{\mu}{2\pi T}\right)  \Big],
\label{eq:IamuT}
\end{split}
\end{equation}
where $\zeta(a,b)$ is the Hurwitz zeta function. Taking $T \to 0$ in this expression is subtle, due to the $\mu / T$ appearing in the $\zeta$ functions. The correct limit can be shown to be 
\begin{equation} \label{eq:I_alpha_T_to_0}
\lim_{T \to 0} I_\alpha(\mu,T) = - \left(  \frac{e^{\gamma_\text{E}} \Lambda^2}{4 \pi}\right)^\epsilon \frac{i \mu}{2 \pi} \frac{\Gamma \left( \alpha - \frac{d}{2} \right)}{(4 \pi)^{d/2} \Gamma(\alpha) (1 + d - 2\alpha)} \left[ (i \mu)^{d-2\alpha} - (-i \mu)^{d-2\alpha} \right].
\end{equation}
Let us now attempt to evaluate $I_\alpha(\mu)$ directly at $T = 0$. If we do this first by first performing the $\mathbf{p}$-integral and subsequently the $p_{0}$-integration, the physically reasonable result $I_\alpha(\mu) = \lim_{T \to 0} I_\alpha(\mu, T)$ appears without any special considerations. However, it is very tempting to perform the $p_0$ integration first, in particular for parameter values $\alpha \in \mathbb{N}$, given how convergent results can be dealt with by using the residue theorem, i.e.\ the first integration would involve only linear algebra and differentiation. The benefits of this approach are more apparent with more complex Feynman diagrams with multiple external legs (or external momentum scales), such as
\begin{eqnarray}
\label{eq:5}
\int_p \int_{-\infty}^\infty \frac{\mathrm{d} p_0}{2 \pi} \frac{1}{\left[\left|\mathbf{p}+ \mathbf{k} \right|^2+(p_0+k_0+i\mu)^2 \right]^2 \left[p^2+(p_0+i\mu)^2 \right]}.
\end{eqnarray}
Even with the $I_\alpha(\mu)$ defined in Eq.~\eqref{eq:Iadef}, performing the $p_0$ integral first leads to an easier computation and gives a completely standard evaluation involving $\Gamma$ functions:
\begin{align}
\label{eq:I_alpha_t_naive}
    \quad I_\alpha(\mu) &\underset{p_0\text{-first}}{\longrightarrow} - \left(  \frac{e^{\gamma_\text{E}} \Lambda^2}{4 \pi}\right)^\epsilon \frac{\mu}{\Gamma \left( \frac{1}{2} \right)} \frac{\Gamma \left( \alpha - \frac{1}{2} \right)}{(4 \pi)^{d/2} \Gamma(\alpha) \Gamma \left( \frac{d}{2} \right)(1 + d - 2\alpha)}  \mu^{d-2\alpha} \\ 
    &\neq \lim_{T \to 0} I_\alpha (\mu, T).
\end{align}
In fact, these two expressions only agree for $\alpha = 1$. We thus surprisingly find that when performing the $T = 0$ evaluation in the simplest way, we obtain a result differing from the $T \rightarrow 0$ limit, which, as a physically motivated value, is what one hopes to find. Since this is the order of integration that one often uses within physical calculations at $T = 0$, it is worth understanding the mechanism responsible for the differences in the final expressions when changing the order of integration. Furthermore, it would be useful to have a procedure for calculating the  value relevant for physical results while still performing the $0$ component first, so that one could continue to use the power of the residue theorem. This is what we set forth to do in this paper.

The outline of our paper is as follows. We start in \Sec\ref{sec:limits} by showing in the simple example of $\alpha = 2$ what is different between the two integration orders at $T = 0$ and suggesting a possible way around these differences. In \Sec\ref{sec:two_int_orders} we systematically study these differences for integer $\alpha$, and demonstrate that the suggested additional boundary terms precisely relate the results following from the two integration orders. In \Sec\ref{sec:thermal_origins}, we demonstrate that these additional terms arise naturally from differentiation of the Fermi--Dirac distribution, which does not explicitly appear in the $T = 0$ expressions. In \Sec\ref{sec:remedial_evaluation}, we complete the calculation for non-integer $\alpha$, which arises in the study of higher-loop effects through dimensional regularization. Finally, in \Sec\ref{sec:results} we summarize our main findings, including practical ways to handle these issues, before concluding with a short discussion. Sections \ref{sec:thermal_origins} and \ref{sec:remedial_evaluation} are quite technical, with the non-integer exponents covered in the latter only showing up in typical computations at two-loop order and above. Accordingly,  readers only interested in the main results should feel free to skip these sections.

\section{The Two Integration Orders: Example Case}
\label{sec:limits}
\noindent 
Let us first evaluate the $T = 0$ integral in Eq.~\eqref{eq:Iadef} for $\alpha = 2$ in the two possible orders. 
Let the superscript $t$ (\emph{temporal}) denote the result with the 0-component integrals being performed first (as opposed to the order immediately leading to physical results, with dimensionally regularized \emph{spatial} integrals taking place first).  We assume here that $\mu > 0$. By using the residue theorem and closing the semicircle contour from the positive half plane, we  obtain
\begin{equation} \label{eq:I2naive}
    \begin{split}
        I_2^t(\mu) &= \int_p \theta(p-\mu) \int_{-\infty}^\infty \frac{\mathrm{d}p_0}{2\pi} \frac{1}{[(p_0+i\mu+ip)(p_0+i\mu-ip)]^2}\\
        &= i \int_p \theta(p-\mu) \left[\frac{\mathrm{d}}{\mathrm{d}p_0} \frac{1}{(p_0+i \mu + ip)^2}\right]_{p_0 \rightarrow -i\mu + ip}\\
        &= -\frac{ \mu^{d-3}}{2 (d-3) (4 \pi)^\frac{d}{2} \Gamma \left( \frac{d}{2} \right)}\left(  \frac{e^{\gamma_\text{E}} \Lambda^2}{4 \pi}\right)^\epsilon,
    \end{split}
\end{equation}
where the step function $\theta$ appears as one observes which poles are enclosed by the integration contour. Additionally, should the outermost integral diverge at large loop momenta (e.g. for $\epsilon < 0$ for $\alpha = 2$ and in general for $\alpha = 1$), we can decompose the step function $\theta(p - \mu) = 1 - \theta(\mu - p)$. This allows us to discard the former part, a scale-free (``vacuum'') integral of the form $\int_p p^\beta$, which vanishes in dimensional regularization. However, if we perform the spatial integral first, as in Eq.~\eqref{eq:Iadef}, we find 
\begin{equation} \label{eq:I2correct}
\begin{split}
I_2 (\mu) &=\left(  \frac{e^{\gamma_\text{E}} \Lambda^2}{4 \pi}\right)^\epsilon\frac{1}{ (4 \pi)^\frac{d}{2} \Gamma \left( \frac{d}{2} \right)} \int_{-\infty}^\infty \frac{\mathrm{d}p_0}{2 \pi} \int_{-\infty}^\infty \mathrm{d}p \frac{(p^2)^\frac{d-1}{2}}{\left[(p-\mu+ip_0)(p+\mu-ip_0)\right]^2}\\
&=-\left(  \frac{e^{\gamma_\text{E}} \Lambda^2}{4 \pi}\right)^\epsilon \frac{i }{ (4 \pi)^\frac{d}{2} \Gamma \left( \frac{d}{2} \right)} \int_{-\infty}^\infty \mathrm{d}p_0 \,\text{sgn}(p_0)  \frac{\mathrm{d}}{\mathrm{d}p} \left[\frac{p^{d-1}}{(p+\mu-ip_0)^2} \right]_{p \rightarrow \mu- ip_0} \\
&= -\frac{(d-2)  \mu^{d-3}}{2(d-3)(4\pi)^\frac{d}{2} \Gamma \left( \frac{d}{2} \right)} \left(  \frac{e^{\gamma_\text{E}} \Lambda^2}{4 \pi}\right)^\epsilon,
\end{split}
\end{equation}
which is a different result. If we examine the integrand here we can gain some insight into why these results are distinct. In particular:
\begin{equation}
\int_p \int_{-\infty}^{\infty} \frac{\mathrm{d} p_0}{2\pi} \frac{1}{|(p_0+i \mu)^2 + p^2|^\alpha} = \infty,
\end{equation}
with a divergence occurring at $p_0 = 0$, $p = \mu$ for $\alpha \geq 1$. Because of this divergence, Fubini's theorem does not apply, and we should not necessarily expect that swapping the integration order should result in the same answer.

Let us remedy this divergence by splitting the problematic point at $p = \mu$ into two parts, so that the integral for $I_2^t$ resembles an integral over two copies of the integrands in $I_1^t$. To  the this end, we consider an integral reminiscent of Eq.~\eqref{eq:5}, with both propagators having exponent 1. By writing  $q \equiv |\mathbf{p}+\mathbf{k}| $ and $k_0 = 0$, we can simplify the expression such that
\begin{align}
\int_{-\infty}^\infty \frac{\mathrm{d}p_0}{2\pi} \frac{1}{(p_0+i\mu+ip)(p_0+i\mu-ip)(p_0+i\mu+iq)(p_0+i\mu-iq)} = \frac{1}{q-p} \left [- \frac{\theta(q - \mu) }{2q} + \frac{\theta(p - \mu)}{2p} \right ] \frac{1}{p+q}.
\end{align}
Taking the limit $q \to p$, we then find the following expression, which can be rewritten in a similar form to the intermediate steps in Eq.~\eqref{eq:I2naive} above:
\begin{equation} \label{eq:polesplit}
\begin{split}
\lim_{q \to p} \frac{1}{q-p} \left [- \frac{\theta(q - \mu) }{2q} + \frac{\theta(p - \mu)}{2p} \right ] \frac{1}{p+q} 
&= \frac{1}{2p} \, \frac{\mathrm{d}}{\mathrm{d} p} \left [- \frac{\theta(p - \mu) }{2p} \right ] \\
&=  i \left[\frac{\mathrm{d}}{\mathrm{d}p_0} \frac{\theta( -i p_0 ) }{(p_0+i \mu + ip)^2}\right]_{p_0 \rightarrow -i\mu + ip} . 
\end{split}
\end{equation}
The difference from Eq.~\eqref{eq:I2naive} is seen in the term which differentiates the theta function; in fact, these additional boundary terms exactly account for the difference between $I_2^t(\mu)$ and $I_2(\mu)$. Such an augmentation to the residue theorem would be a convenient solution, as it is simple to implement in practical situations. However, so far it is merely an observation, and we must verify if it works in general, and if so, why. We shall address this in further sections.

For the time being, however, let us note that this pole-merging analysis can be easily generalized to integer-valued exponents $m \in \mathbb{N}$. Generalizing \eqref{eq:polesplit}  would then lead to expressions of the form 
\begin{equation} \label{eq:diff}
\begin{split}
\int_\mathbb{R} \frac{\mathrm{d}z}{2 \pi} \frac{1}{[(z+i\mu)^2+p^2]^m} &\overset{?}{=}  \left.\frac{i}{(m-1)!} \frac{d^{m-1}}{\mathrm{d}z^{m-1}} \frac{\theta(-iz)}{(z+i\mu + ip)^m}\right|_{z= -i \mu +ip} \\
&=\frac{(-1)^{m+1} }{\Gamma (m)}\sum_{k=0}^{m-1}  \frac{\Gamma (m+k)}{\Gamma (k+1) \Gamma (m-k)} \frac{(-1)^k  \theta^{(m-1-k)} (p-\mu) }{(2p)^{m+k}},
\end{split}
\end{equation}
whereas simply applying the residue theorem would produce only the final term with $k = m - 1$. We will presently demonstrate that these additional terms exactly account for the difference between $I_m^t(\mu)$ and $I_m(\mu)$ for general integer values for m. We emphasize that for $m > 1$, all of these correction terms arise from the single point $p = \mu$, where the denominator of the integrand contains a real pole of order $m$ at $z = 0$, and which we previously identified as a possible problem due to the breakdown of Fubini's theorem. This suggests that our application of the residue theorem does not recognize these delta distributions at the boundary $p = \mu$, associated with higher-order poles and the finite temperature result, but rather returns only a smooth result there.

Let us finally make one final remark here about an alternative formulation of the above series. One may also arrive at the series in Eq.~\eqref{eq:diff} for integer-valued $m$ by considering rewriting the original integral in terms of spatial differentiation of $I^t_1(\mu)$ as follows
\begin{equation}\label{eq:IBP}
\int_p \int_{-\infty}^\infty \mathrm{d}p_0 \frac{1}{[(p_0+i\mu)^2+p^2]^m} = \frac{(-1)^{m-1}}{(m-1)!} \int_p \left( \frac{\mathrm{d}}{\mathrm{d} p^2} \right)^{m-1} \int_{-\infty}^\infty \mathrm{d}p_0 \frac{1}{(p_0+i\mu)^2+p^2}.
\end{equation}
This strategy essentially allows one to consider the computation as an integration-by-parts (IBP) problem \cite{Chetyrkin:1981qh} for the 0-component integral of $I_1(\mu)$, where the residue theorem produces the sought-after/physically motivated result. Given the simplicity of the IBP procedure and, particularly since the development of the Laporta algorithm \cite{Laporta:2000dsw}, the widespread applicability to vacuum quantum field theory \cite{Czakon:2004bu,Anastasiou:2002yz,Anastasiou:2003ds,Schroder:2005hy,Chetyrkin:2005ia} and some thermal problems \cite{Nishimura:2012ee}, the result seen in Eq.~\eqref{eq:diff} is a promising step to applying the method to finite-density field theory. To confirm this, let us study the results found from different integration orders in the following section. 

\section{\texorpdfstring{General Integer Exponents $\alpha$}{General Integer Exponents α}}
\label{sec:two_int_orders}
\noindent
We have already presented the result from the standard residue evaluation for integer-valued $\alpha$, corresponding to $I^t_{\alpha}(\mu)$, in Eq.~\eqref{eq:I_alpha_t_naive}. 
Since the evaluation of $I_\alpha(\mu)$ is not terrifyingly lengthy, we will give a rather detailed description below which we can then compare with $I^t_\alpha(\mu)$. Since we have evaluated $I^t_\alpha(\mu)$ using the residue theorem, we are presently limited to $\alpha \in \mathbb{N}$ for this comparison. However, the spatial integral can just as easily be performed for any $\alpha \in \mathbb{R}$, which will show similarities to the evaluation of the general case in further sections. These non-integer exponents do not appear in standard computations \emph{up to two loops}, but do contribute at higher orders. Additionally, they yield insight on the overall behavior and the role of dimensional regularization in this puzzle. 

\subsection{Spatial-Temporal Integration Order}
\noindent
For the evaluation of $I_\alpha(\mu)$ with real-valued $\alpha$, the toolkit of the computation moves from the residue theorem to analytic continuation of Euler's Beta functions.

We proceed in some generality. Let us define four real and non-zero parameters $\{\beta, \gamma, y, c\}$, and consider the following analytic continuation of the Euler Beta function 
\begin{equation}
\begin{split}
    \int_0^\infty \frac{\mathrm{d}x x^\beta}{(x+y+ic)^\gamma}
    &=\left[y+ic \right]^{\beta+1-\gamma}\int_0^{\frac{y-ic}{c^2+y^2} \infty}  \frac{\mathrm{d}z z^\beta}{[z+1]^\gamma},
    \end{split}
\end{equation}
where the last step in particular requires $c \neq 0$ or $y > 0$. The standard Beta function corresponds to taking $y + i c \mapsto 1$, and so the above integral corresponds to integrating at a more general angle in the complex plane. For the real-valued integrals with $c=0$ there are many other applicable methods for evaluating this expression; here we are only concerned with $c\neq 0$, since $c$ is associated with a chemical potential. 

The integral can be associated to a closed pizza slice contour (of infinite radius), avoiding the pole at $z=-1$. We shall assume $\beta$ and $\gamma$ to be such that the integral along the arc of the contour vanishes ($\beta-\gamma+1 < 0$), which leads to  
\begin{equation}
\begin{split}
 -\left[y+ic \right]^{\beta+1-\gamma}\int_{\infty}^{0}  \frac{\mathrm{d}z z^\beta}{[z+1]^\gamma} 
&=\left[y+ic \right]^{\beta+1-\gamma} \frac{\Gamma \left( \beta+1 \right)\Gamma \left(\gamma- \beta-1 \right)}{\Gamma (\gamma)}.
\end{split}
\end{equation}
With this result in mind, let us return to spatial integral of interest, where we can factor out the volume of the $(d-1)$-sphere $\Omega_d$, and change variables in the radial integral to obtain
\begin{equation}
\begin{split}
\Omega_d \int_0^\infty \mathrm{d}p \frac{p^{d-1}}{[p^2+(p_0+i\mu)^2]^\alpha} &= \frac{\Omega_d}{2} \int_0^\infty \frac{\mathrm{d}y y^{\frac{d}{2}-1}}{[y+ (p_0+ i \mu)^2]^\alpha} \\
&=\frac{\pi^\frac{d}{2} \Gamma \left(\alpha- \frac{d}{2} \right)}{\Gamma (\alpha)}[(p_0+i\mu)^2]^\frac{d-2\alpha}{2}.
\end{split}
\end{equation}

The remaining computation involves using this result in the $p_0$ integral, which we must break up in two pieces. Accordingly, we have
\begin{equation}
\begin{split}
\int_0^\infty \mathrm{d}p_0 \left[ (p_0+i\mu)^2 \right]^\frac{d-2 \alpha}{2}+ \int_0^{\infty} \mathrm{d}p_0 \left[ (-p_0+i\mu)^2 \right]^\frac{d-2 \alpha}{2} &= 2 \Re \left\{\int_0^\infty \mathrm{d}p_0 \left[ (p_0+i\mu)^2 \right]^\frac{d-2 \alpha}{2} \right\}\\
&= -2 \Re \left[\frac{ (i \mu)^{d+1-2\alpha}}{d+1-2 \alpha}  \right].
\end{split}
\end{equation}
Upon combining the two intermediate results, we find the following expression
\begin{equation}
I_{\alpha}(\mu)=- \left(\frac{e^{\gamma_\text{E}} \Lambda^2}{4 \pi}\right)^\epsilon \frac{\ \Gamma \left(\alpha-\frac{d}{2} \right)\cos \left[\frac{\pi}{2} (d+1-2 \alpha) \right]}{\pi (4 \pi)^\frac{d}{2} \Gamma (\alpha)(d+1-2 \alpha)} \mu^{d+1-2 \alpha}.
\end{equation}
By applying trigonometric algebra in combination with Euler's reflection formula, we can simplify this expression further. The more compact expression is given in terms of $\Gamma$ functions and reads 
\begin{equation}
\label{eq:zero-result}
I_\alpha (\mu) = -\left(  \frac{e^{\gamma_\text{E}} \Lambda^2}{4 \pi}\right)^\epsilon\frac{ 1}{ (4 \pi)^\frac{d}{2} \Gamma (\alpha) \Gamma \left(\frac{d}{2}+1-\alpha \right)}\frac{\mu^{d+1-2 \alpha}}{(d+1-2 \alpha)}.
\end{equation}
This result agrees with the $T \to 0$ limit of the $T > 0$ expression from \cite{Vuorinen:2003fs}, given in  Eq.~\eqref{eq:I_alpha_T_to_0}. This justifies our choice of the  integration order in Eq.~\eqref{eq:Iadef}, as well as our choice to refer to this integration order as the physically motivated one. 

We also emphasize here that the result in Eq.~\eqref{eq:zero-result} can be used to confirm the viability of the IBP strategy used above. We can find an explicit mapping from $\alpha \mapsto \alpha+1$ by writing 
 \begin{equation*}
\begin{split}
 \frac{1}{2\alpha}\int_{-\infty}^\infty \mathrm{d}p_0 \int_0^\infty \mathrm{d}p p^{d-2} \frac{\mathrm{d}}{\mathrm{d}p}  \frac{1}{[(p_0+i\mu)^2+p^2]^\alpha} &\mapsto-\frac{d-2}{2\alpha} \int_{-\infty}^\infty \mathrm{d}p_0 \int_0^\infty \mathrm{d}p    \frac{p^{d-3}}{[(p_0+i\mu)^2+p^2]^\alpha}\\
&=-\frac{ \pi \Gamma \left( \frac{d}{2} \right)  }{  \Gamma (\alpha+1) \Gamma \left(\frac{d}{2}-1-\alpha \right)}\frac{\mu^{d-1-2 \alpha}}{(d-1-2 \alpha)},
 \end{split}
 \end{equation*}
which indeed implies that unit step in the exponent takes place via operation $-\frac{1}{\alpha} \frac{\mathrm{d}}{d p^2}$ inside the integrand of $I_\alpha (\mu)$. 

\subsection{Summary of the Differences}
\noindent
Let us now see how our hypothesis for amending $I^t_\alpha(\mu)$ with boundary terms compares to the value computed above for $I_\alpha(\mu)$ in the case of integer $\alpha$. To this end, let us define $I^{t,\text{new}}_\alpha(\mu)$ to be the amended result, using Eq.~\eqref{eq:diff} in place of the naive residue result. Applying the summation seen in Eq.~\eqref{eq:diff} and integrating derivatives of $\delta$ functions by parts, we find
\begin{equation}
\begin{split}
\int_0^\infty \mathrm{d}p p^{d-1} &\int_{-\infty}^\infty \frac{\mathrm{d}p_0}{2 \pi i} \frac{1}{\left[p^2+(p_0+i\mu)^2\right]^\alpha} 
= i\frac{(-1)^{2\alpha}  \Gamma (2\alpha-1)}{2^{2\alpha-1}\Gamma^2 (\alpha)}\frac{\mu^{d+1-2\alpha}}{d+1-2\alpha}\\
&+ \theta(\alpha-2)\frac{(-1)^\alpha i }{\Gamma (\alpha)}\sum_{k=0}^{\alpha-2}  \frac{(-1)^k\Gamma (\alpha+k)}{2^{\alpha+k}\Gamma (k+1) \Gamma (\alpha-k)} (-1)^{\alpha-2-k} \int_0^\infty dp  \delta(p-\mu) \left( \frac{\mathrm{d}}{\mathrm{d}p} \right)^{\alpha-2-k} p^{d-\alpha-k-1}.
\end{split}
\end{equation}
Here the first row stands for the naive residue value, while the second describes the contribution from the added differentiated step functions, with nonzero contributions from $\alpha \geq 2$ as indicated by the step function. For $\alpha=1$, these terms vanish and indeed $I_1^{t,\mathrm{new}} = I_1^t = I_1$. Multiplying this intermediate result by the necessary factor, and performing the remaining integrals we find 
\begin{equation}
\begin{split}
I_\alpha^{t,\text{new}} (\mu)=& I^t_\alpha(\mu) 
-\left(  \frac{e^{\gamma_\text{E}} \Lambda^2}{4 \pi}\right)^\epsilon  \theta(\alpha-2)\mu^{d+1-2\alpha} \frac{2(-1)^\alpha }{\Gamma (\alpha)\Gamma \left(\frac{d}{2} \right)(4\pi)^\frac{d}{2}}  \sum_{k=0}^{\alpha-2}  \frac{(-1)^k\Gamma (\alpha+k)}{2^{\alpha+k}\Gamma (k+1) \Gamma (m-k)} \frac{\Gamma (2\alpha-d-1)}{\Gamma(\alpha+k+1-d)} .
\end{split}
\end{equation}

The first three integer values of $\alpha$ can be simplified to give
\begin{eqnarray}
I_1^{t,\text{new}}(\mu) &=&  
I_1^t(\mu)\\
I_2^{t,\text{new}} (\mu) &=& -\left(  \frac{e^{\gamma_\text{E}} \Lambda^2}{4 \pi}\right)^\epsilon \frac{1 }{2 (4 \pi)^\frac{d}{2} \Gamma \left(\frac{d}{2} \right)}\frac{(d-2)\mu^{d-3}}{d-3}
\end{eqnarray}
and
\begin{equation}
I_3^{t,\text{new}} (\mu)  
=-\left(  \frac{e^{\gamma_\text{E}} \Lambda^2}{4 \pi}\right)^\epsilon\frac{1 }{8 (4 \pi)^\frac{d}{2} \Gamma \left(\frac{d}{2} \right)}\frac{\mu^{d-5}} {d-5}(d-2)(d-4),
\end{equation}
which agree with the results for $I_\alpha (\mu)$ in Eq.~\eqref{eq:zero-result}. One can similarly verify agreement between $I_\alpha$ and $I_\alpha^{t,\text{new}}$ for all integer $\alpha$ up to any given finite value, although the process becomes increasingly tedious, involving an increasing number of correction terms. Given this, we note that our originally somewhat intuitively motivated prediction also agrees with the physical finite-temperature limit.  What remains is to understand why this works, or rather what causes the need for this kind of treatment, which we address next. 

\section{Thermal Origin of the Boundary Terms}
\label{sec:thermal_origins}
\noindent
While considering the zero-temperature integral, we have already recognized the difference in how the residue theorem treats (or rather, doesn't treat) the boundary associated to the pole from how dimensional regularization handles it. The pieces necessary to match the results appear as boundary terms associated with the zero-component integral. Their specific form is as derivatives of step functions, which is the zero-temperature limit of the Fermi--Dirac distribution function. Given both this and the fact that our physically motivated result $I_\alpha$ arises as a zero-temperature limit, one could expect to find a better explanation for the mechanism behind this difference by examining the full expression at finite temperature. 

\begin{figure}[t!] 
\centering 
\includegraphics[width=1.0\textwidth]{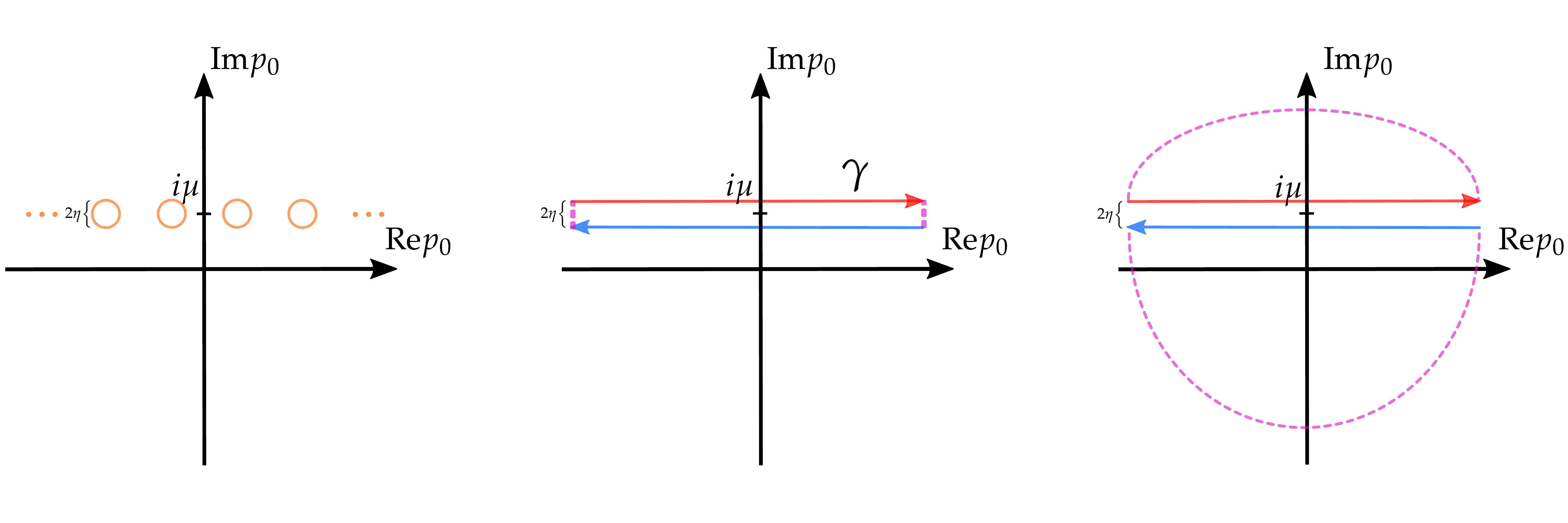}
\caption{A depiction of the integration contours associated with fermionic Matsubara sums. The dashed lines indicate contributions that are taken to vanish.}
\label{fig:matsubaracontour}
\end{figure}

At finite temperature, the $p_0$-integral in Eq.~\eqref{eq:Iadef} is replaced by a frequency sum over fermionic Matsubara frequencies, as in Eq.~\eqref{eq:IamuT}.  
The sum is often 
re-cast into a contour integral using
\begin{equation}
\begin{split}
\label{eq:fermi_dirac_contour}
I_\alpha(\mu, T) = \int_p \left[\int_{-\infty+i\mu+i\eta}^{\infty+i \mu+i\eta }+\int_{\infty+i\mu-i\eta}^{-\infty+i \mu -i\eta } \right] \frac{\mathrm{d}p_0}{2 \pi}  \frac{1}{[p^2+p_0^2]^\alpha} n_F \left[i\beta\left(p_0-i\mu \right) \right]
\end{split}
\end{equation}
with $\beta = 1/T$, and where $\eta > 0$ is a small regulator to avoid the poles in the complex distribution function, and $n_F(x) = 1/[\exp(x) + 1]$ is the Fermi--Dirac distribution function. The box contour $\gamma$ used in this expression is defined in Fig.~\ref{fig:matsubaracontour}. 

This  definition has the benefit of being very regular, and can be easily shown through direct computation to yield the same result in either order of integration. Thus, it is well-motivated---given our earlier considerations---to study the small-temperature behavior of this expression and see how the boundary terms are generated. 

We can write the Fermi--Dirac occupation function restricted to the non-vanishing sides of the contour as
\begin{equation} \label{eq:fd_distribution}
\frac{1}{e^{-\beta\left(\Im[p_0]-\mu\right)} \left[ \cos \left(\beta\Re[p_0] \right)+i \sin \left(\beta\Re[p_0] \right) \right]+1} = \frac{1}{ e^{\pm\beta\eta} \left[ \cos \left(\beta\Re[p_0]  \right)+i \sin \left(\beta\Re[p_0]  \right) \right]+1}.
\end{equation}
To examine the behavior near zero-temperature, we consider the hierarchies $0< T \ll \eta$ and $\eta < \mu$ and note that the trigonometric part is non-vanishing for all possible values of $p_0$. Depending on the sign in the exponential of the right-hand side of Eq.~\eqref{eq:fd_distribution}, the zero-temperature limit yields a different answer, which leads to the standard limit for the distribution function (here written as a complex generalization, as $p_0$ is not a true component of a four-momentum, but rather a complex-valued integration variable), 
\begin{equation} \label{eq:fdlimit}
n_F\left[i\beta\left(p_0-i\mu\right)\right] \underset{T \rightarrow 0}{\longrightarrow} \theta \left(\Im[p_0]-\mu \right).
\end{equation}
The limit is ambiguous at exactly $\Im[p_0]-\mu = 0$, where we obtain a highly oscillatory term with the already familiar set of real-part divergences $\lbrace 
1/(e^{i\beta\Re[p_0]}+1) \rbrace_{p_0}$
and which motivates the original box integral, leading back to the summation formula in Eq.~\eqref{eq:IamuT}.

Assuming the limit \eqref{eq:fdlimit}, the box contour would indeed yield
\begin{equation}
\label{eq:secondbox}
\begin{split}
\int_p \oint_{\gamma} \frac{\mathrm{d}p_0}{2 \pi}\frac{1}{[p^2+p_0^2]^\alpha} n_F \left[i\beta\left(p_0-i\mu\right)\right] &\longrightarrow \int_p\int_{-\infty+i\mu+i\eta}^{\infty+i \mu+i\eta } \frac{\mathrm{d}p_0}{2\pi} \frac{\theta(\Im[p_0]-\mu)}{[p^2+ p_0^2]^\alpha} \\
&= \int_p\int_{-\infty}^{\infty} \frac{\mathrm{d}p_0}{2\pi} \frac{1}{\left[p^2+ \left( p_0+i\mu+i\eta \right)^2 \right]^\alpha}.
\end{split}
\end{equation}
which agrees with our initial zero-temperature integral (apart from the $\eta$ shift in the propagator). Of course, this expression comes with the caveat that we exchanged the $T \to 0$ limit and the integrals. Note that in Eq.~\eqref{eq:secondbox} above, exchanging the $T \to 0$ limit and the integral rejects the lower part of $\gamma$ in the complex plane, and allows us to rewrite the integral in a way such that the occupation function is no longer seen explicitly.
This indeed appears to be both the reason why $I_\alpha^t$ does not describe the physically motivated zero-temperature limit, as well as the origin of the differentiation formula. 

Let us now compute this $T \to 0$ more carefully, starting with $\alpha \in \mathbb{N}$. After splitting $\gamma$ into two line integrals parallel to real axis, we note that each of the two $p_0$ integrals can be computed via the residue theorem in the complex plane via \emph{convergent} semicircles, closing above or below depending on their location relative to the line $p_0 = i \mu$ (see the final panel of Fig.~\ref{fig:matsubaracontour}).
This procedure gives rise to a second step function independent of the one obtained from the Fermi-Dirac occupation function, related to whether the poles in the propagator are within the semicircles or not. Explicitly, we find for the upper horizontal line 
\begin{equation}
\begin{split}
\label{eq:upperline}
\int_p \int_{-\infty+i\mu+i\eta}^{\infty+i \mu+i\eta } \frac{\mathrm{d}p_0}{2 \pi}  \frac{1}{[p^2+p_0^2]^\alpha} n_F \left[i\beta\left(p_0-i\mu\right)\right]= \frac{i}{(\alpha-1)!} \int_p \theta(p-\mu) \frac{\mathrm{d}^{\alpha-1}}{\mathrm{d}p_0^{\alpha-1}} \left[ \frac{n_F \left[i\beta\left(p_0-i\mu\right) \right] }{(p_0+ip)^\alpha} \right]_{p_{0} \rightarrow i p}
\end{split}
\end{equation}
and for the lower horizontal line 
\begin{equation}
\begin{split}
\label{eq:lowerline}
\int_p \int_{-\infty+i\mu-i\eta}^{\infty+i \mu-i\eta } \frac{\mathrm{d}p_0}{2 \pi}  \frac{1}{[p^2+p_0^2]^\alpha} n_F \left[i\beta\left(p_0-i\mu \right) \right]=&\frac{i}{(\alpha-1)!} \int_p \frac{\mathrm{d}^{\alpha-1}}{\mathrm{d}p_0^{\alpha-1}} \left[\frac{n_F \left[i\beta\left(p_0-i\mu\right) \right] }{(p_0-ip)^\alpha}\right]_{p_{0} \rightarrow -i p}\\
&+\frac{i}{(\alpha-1)!} \int_p \theta(\mu-p) \frac{\mathrm{d}^{\alpha-1}}{\mathrm{d}p_0^{\alpha-1}} \left[ \frac{n_F \left[i\beta\left(p_0-i\mu\right) \right] }{(p_0-ip)^\alpha} \right]_{p_{0} \rightarrow i p}.
\end{split}
\end{equation}

To demonstrate the effect of these formulas more explicitly, let us first consider  $\alpha = 2$. With the knowledge of how the occupation function tends towards the step function according to Eq.~\eqref{eq:fdlimit}, it is quite easy to visualize that the $\delta$-function contributions manifest from this expression as expected. A more rigorous analysis involves differentiating with respect to $p_0$ before taking the limit of interest. For this purpose, we note
\begin{equation}
\frac{\mathrm{d}}{\mathrm{d}x} n_F(x) = -\frac{e^x}{(e^x+1)^2} =n_F(x) [n_F(x)-1].
\end{equation}
This implies that the integrands in Eqs.~\eqref{eq:upperline}-\eqref{eq:lowerline} are proportional to  the Fermi-Dirac distribution, which indeed allows us to ignore the lower line integral in the small-$T$ limit [as it is $O \left( e^{-\eta/T}\right)$]. For the upper line, we can isolate the correction term arising from the derivative of the distribution function as 
\begin{equation}
\begin{split}
-\beta \,n_F'\left[i\beta \left(p_0-i\mu\right)\right] &=- \beta\, n_F \left[ \beta\left(\mu-p\right) \right] \left\{ n_F \left[ \beta\left(\mu-p\right) \right] -1 \right\}.\\
\end{split}
\end{equation}
Setting $x = \mu-p$, we can rewrite this expression as 
\begin{equation}
-\beta n_\text{F} \left(\beta x \right) \left[n_\text{F} \left(\beta x\right)-1 \right]= 
 \frac{1}{4 T \cosh^2 \left( \beta x/2\right)} 
\equiv \delta_{2T}(x),
\end{equation}
where we recognize a family of nascent delta functions. Thus, returning to the original expression, we can in good faith write the sought-after limit
\begin{equation}
\underset{T \rightarrow 0}{\lim} \left\{- \frac{1}{T} n_F \left( \frac{\mu-p}{T} \right) \left[n_F \left( \frac{\mu-p}{T} \right)-1 \right] \right\} = \delta(\mu-p).
\end{equation} 

For the cases $\alpha > 2$, we recognize that additional boundary contributions would indeed be seen as derivatives of nascent delta functions $\delta_{2T}^{(j)}(\mu-p)$ with $1 \leq j \leq \alpha-2$. The additional term with most derivatives acting on the occupation function would always correspond to
\begin{equation}
  -\frac{1}{(\alpha-1)!(2p)^\alpha}  \left( \frac{1}{T} \right)^{\alpha-2} \delta_{2T}^{(\alpha-2)}(\mu-p),
\end{equation}
where the structure can be easily associated to the corresponding final term in the sum of Eq.~\eqref{eq:diff} after careful applications of IBP followed by the $T \to 0$ limit.

The above analysis can be applied to the iterated differentiation seen in Eq.~\eqref{eq:upperline}. Doing so, we find the elements of the sum of Eq.~\eqref{eq:diff}, and, as suggested earlier, observe the same difference between  Eq.~\eqref{eq:zero-result}  and  Eq.~\eqref{eq:I_alpha_t_naive} in the zero-temperature limit. The temperature acts as an additional cutoff of the quantum field theory, but from the point of view of complex analysis, it mollifies the distribution function, causes the integral along the arc of the integration contour in Fig.~\ref{fig:matsubaracontour} to vanish, and allows us to discard exponentially decaying contributions $O\left( e^{-\eta/T} \right)$. The parameter $\eta$, present also in the contour formulation at zero temperature, serves a similar purpose, but is somewhat less transparent.

We also see that the result is consistent if the integration order is reversed: Taking the leading-order low-temperature limit and computing the spatial integral, we have, up to an overall constant multiple,  
\begin{equation}
\begin{split}
\int_{-\infty+i\mu+i\eta}^{\infty+i\mu+i\eta} \frac{\mathrm{d}p_0}{2\pi} \left(p_0^2 \right)^{\frac{d}{2}-\alpha} n_F\left[i\beta \left(p_0-i\mu\right)\right]
&=\int_{-\infty+i\mu+i\eta}^{\infty+i\mu+i\eta} \frac{\mathrm{d}p_0}{2\pi} \left(p_0^2 \right)^{\frac{d}{2}-\alpha} \left(1+\left\{ n_F\left[i\beta \left(p_0-i\mu\right)\right]-1  \right\}\right)\\
&= \int_{-\infty+i\mu+i\eta}^{\infty+i\mu+i\eta} \frac{\mathrm{d}p_0}{2\pi} \left(p_0^2 \right)^{\frac{d}{2}-\alpha} \left\{1+O \left( e^{-\beta\eta} \right)\right\}.\\
\end{split}
\end{equation}
Thus, the zero-temperature limit matches with  $I_\alpha(\mu)$ performed at exactly zero temperature. This is one last reassurance that the earlier results  $I_\alpha$ and $I_\alpha^{t,\text{new}}$ are indeed the ones relevant for physics,  while also explicitly demonstrating the finite-temperature origin of the discrepancy and the interplay between the zero-temperature limit and dimensional regularization. 

\section{\texorpdfstring{Non-Integer Exponents $\alpha$}{Non-Integer Exponents α}}
\label{sec:remedial_evaluation}
\noindent
Let us now generalize the previous section to non-integer $\alpha \in \mathbb{R}$. This is relevant for example when considering terms involving $I_\alpha(\mu)$ for $\alpha\in \mathbb{Q}[d]$ multiplying divergent expressions \cite{Vuorinen:2003fs}. Since Eq.~\eqref{eq:zero-result} (which agrees with the careful zero-temperature limit) can be extended to arbitrary values of $\alpha \in \mathbb{R}$, we already have an expression to compare against. Fubini's alone theorem would imply convergence for $\alpha<1$, but for non-integer values of $\alpha$ there are additional contributions which cannot be obtained simply by using the residue theorem and taking (an integer-valued number of) derivatives. Perhaps the most transparent way of studying the missing contributions is by carrying out a careful limit procedure of the $T > 0$ expressions to see how the boundary terms analogous to those seen previously are generated from contour integrals. 

First note that for any positive $\alpha$, we can use IBP (in the dimensionally regularized sense) as in Eq.~\eqref{eq:IBP} to relate the integral to one involving $\alpha \in (0,1]$. In terms of the residue theorem, $\alpha = 1$ is an important limit, being the smallest integer value that allows one to perform the $p_0$-integral by completing the contour as a semicircle (and to apply Jordan's lemma without issues). Any value of the parameter $\alpha \in (0,1)$ is in some sense challenging. Given these potential convergence issues with arc contours, we are interested in finding a box-like contour involving the real axis and a line parallel to it. Such an approach is necessary to compute even the generalized version of the naive result, $I_\alpha^t (\mu)$, which can be associated to a Beta-function integral along the real axis.

The full calculation of the $T \to 0$ limit requires one to consider separately two hierarchies of the radial coordinate $p$, namely, one with $p > \mu$ and the other with $p < \mu$. The former will be easily found to correspond to $I_\alpha^t(\mu)$, while the latter requires more care and will vanish at $\alpha = 1$. 

We begin with the expression for $I_\alpha(\mu, T)$. We first manipulate it to obtain an integrand that vanishes for $\Im(p^0) \to \infty$. To this end, we note the Fermi-Dirac decomposition of unity given by 
\begin{equation}
n_F(x)+n_F(-x) = 1,
\end{equation}
to re-write Eq.~\eqref{eq:fermi_dirac_contour} as
\begin{equation}
\label{eq:box_v2}
I_\alpha(\mu, T)=\int_p \int_{-\infty+i\mu-i\eta}^{\infty+i \mu-i\eta } \frac{\mathrm{d}p_0}{2 \pi}  \frac{1}{[p^2+p_0^2]^\alpha} n_F \left[-i\left(\frac{p_0-i\mu}{T} \right) \right]+ O\left(e^{-\frac{\eta}{T}} \right),
\end{equation}
again imposing the hierarchy $0<T\ll\eta$. We now have a distribution function which tends towards unity everywhere in the complex plane with $\Im[p_0] < \mu$. This  is precisely the region where we intend to complete the box contour, since we recognize that it is not possible to apply the previous (infinite) semi-circular contour arguments.

We now create a contour $\Gamma$ consisting of the real line and the line $p^0 = i \mu - i \eta$. Upon closing it, there is at most a single pole within the new box contour. In particular, for $p > \mu$ there are none, which allows us to move from the complex contour to the real axis and continue by taking $T\to0$. This gives, for the part of $I_\alpha(\mu, T)$ arising from large momenta $p>\mu$, 
       \begin{equation}
           \begin{split}
      I^{p>\mu}_\alpha(\mu, T) =\int_p \theta(p-\mu) \int_{-\infty}^{\infty} \frac{\mathrm{d}p_0}{2 \pi}  \frac{1}{[p^2+p_0^2]^\alpha} n_F \left[-i\left(\frac{p_0-i\mu}{T} \right) \right]
       &\underset{T \rightarrow 0}{\longrightarrow}\int_p \theta(\mu-p) \int_{-\infty}^{\infty} \frac{\mathrm{d}p_0}{2 \pi} \frac{\theta \left(\mu-\Im[p_0] \right)}{[p^2+p_0^2]^\alpha} \\
       &=-\left(  \frac{e^{\gamma_\text{E}} \Lambda^2}{4 \pi}\right)^\epsilon \frac{\mu^{d+1-2 \alpha}}{\sqrt{\pi}(d+1-2\alpha)(4 \pi)^\frac{d}{2} } \frac{\Gamma \left( \alpha-\frac{1}{2}\right)}{\Gamma (\alpha) \Gamma \left(\frac{d}{2} \right)}.
    \end{split}
    \label{eq:theta_p-mu}
\end{equation}
This expression corresponds precisely to $I_\alpha^t(\mu)$ in Eq.~\eqref{eq:I_alpha_t_naive}, as here the theta function keeps the $p$ integral away from the problematic point at $\mu$. 

\begin{figure}[t!] 
\centering 
\includegraphics[width=0.30\textwidth]{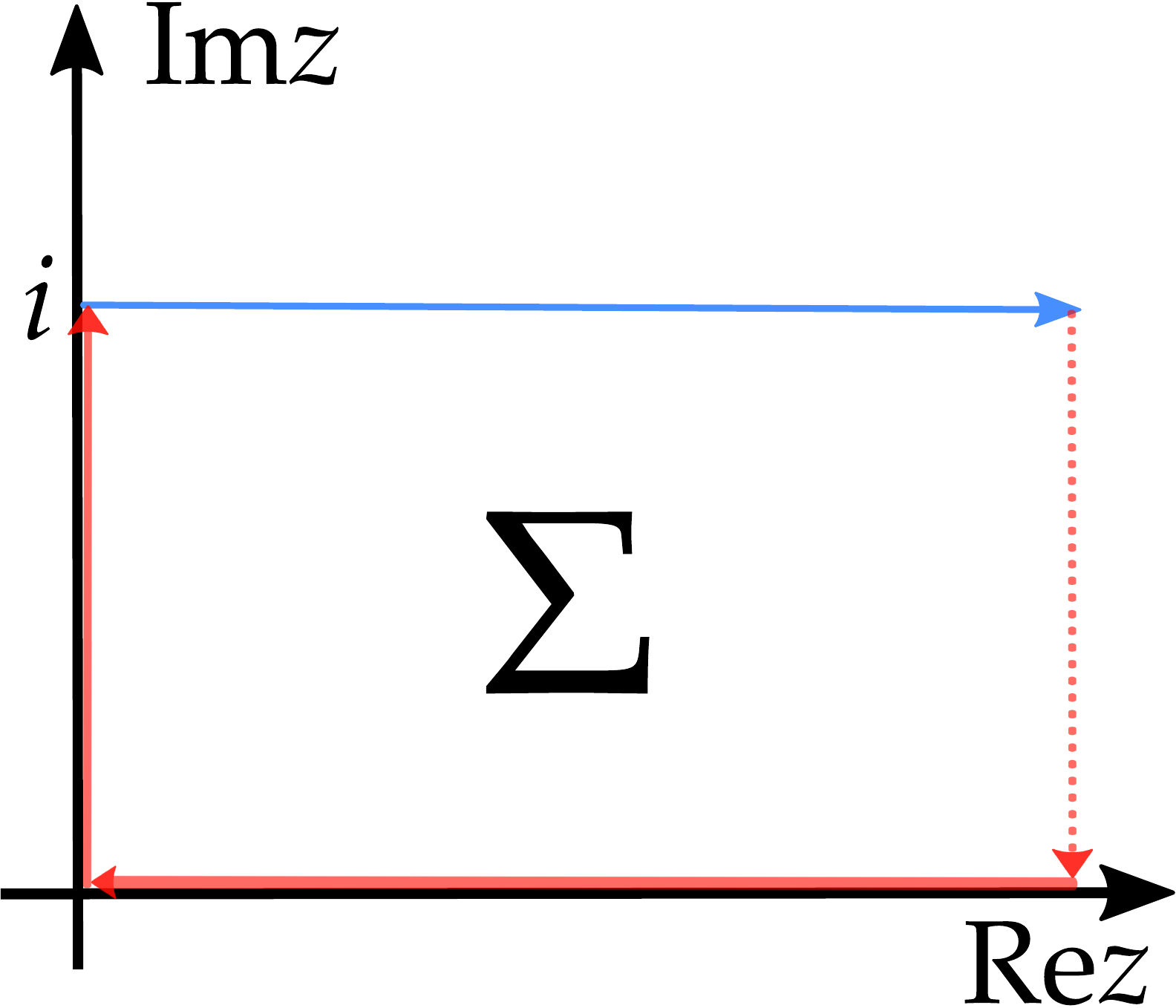}
\caption{Blue color signifies the infinite line segment $(i,\infty+i)$ which constitutes the initial integration domain. Together with the red line segments it forms the closed contour $\Sigma$, inside which the integrand is holomorphic. Note that the integral along the dashed red line at real infinity vanishes.  }
\label{fig:dimensions12}
\end{figure}

In the region of small momenta $p < \mu$ we cannot find an equally nice expression, due to the pole lying within the 
(previous) integration contour. Hence, we must find an alternative way to evaluate the integral. Let us take the $T \to 0$ limit and scale out $\mu$ from the integrands to arrive at
\begin{equation}
\begin{split}
I^{p<\mu}_\alpha(\mu, T) =&\left(  \frac{e^{\gamma_\text{E}} \Lambda^2}{4 \pi}\right)^\epsilon \frac{ \Omega_d \mu^{d+1-2\alpha}}{ (2\pi)^{d+1}} \int_0^\infty \mathrm{d}p p^{d-1} \theta(1-p) \int_{-\infty}^\infty \frac{\mathrm{d}p_0}{[(p_0+i)^2+p^2]^\alpha},
\end{split}
\end{equation}
which we note is equivalent (up to a factor of two) to integrating over the positive real axis for $p_0$ and taking the real part of the expression. Herein we consider the lower limit of this integral over $p_0$ to be regulated by a small positive infinitesimal to avoid a non-integrable pole occurring at $p = 1$ for $\alpha > 1$. Furthermore, let us focus on this modified integral without the trivial overall multipliers 
\begin{equation}
\Re \left[\int_0^1 dp p^{d-1} \int_{i}^{\infty+i} \frac{d z}{[z^2+p^2]^\alpha} \right].
\end{equation}
We can recognize that the biggest computational (regulatory) challenge arises from integration region in which both $p, |z| \sim 1$. Inspired by Cauchy's integral theorem, we aim to extract more approachable line integrals by re-writing the innermost integral in terms of a closed box contour, $\Sigma$, shown in Fig.~\ref{fig:dimensions12}. The contour $\Sigma$ completes the existing line integral, with none of the poles being contained \emph{inside} the region bounded by $\Sigma$, so that
\begin{equation}
    \oint_\Sigma \frac{\mathrm{d}z}{[z^2+p^2]^\alpha} = 0.
\label{eq:sigma_vanish}
\end{equation}
This formulation isolates all the troublesome elements on the line integral on the left-hand side of $\Sigma$, while the right-hand side line integral can be assumed to vanish for all $\alpha >0$ given that $z \in (\infty + i, \infty)$.

The contribution from the real axis (with its direction as depicted in Fig.~\ref{fig:dimensions12} above) is evaluated with ease, yielding
\begin{equation}
\begin{split}
 I_{\Sigma, \text{lower}} = -\Re \left[ \int_0^1 \mathrm{d}p p^{d-1} \int_0^{\infty} \frac{\mathrm{d} z}{[z^2+p^2]^\alpha}  \right] = -\frac{\sqrt{\pi}}{2(d+1-2\alpha) } \frac{\Gamma \left( \alpha-\frac{1}{2}\right)}{\Gamma (\alpha) }. 
\end{split}
\label{eq:horiz_int}
\end{equation}
As the $p$-integral is cut off from above instead of below, the integral now converges for $(d+1)/2> \alpha >1/2$. 
Once the numerical coefficients are added back, we see that this contribution equals $I_\alpha^t(\mu)$ as well, and so will cancel the $p>\mu$ contribution computed above, once one uses the fact that the integral over $\Sigma$ vanishes [Eq. \eqref{eq:sigma_vanish}].

The remaining line integral along the imaginary axis can be split between the hierarchies $p >|z| $ and $|z| > p$, yielding  
\begin{equation}
\label{eq:lineintegralhierarchies}
\begin{split}
I_{\Sigma, \text{left}} = \Re \left[  \int_0^1 \mathrm{d}p p^{d-1} \int_0^i \frac{\mathrm{d}z}{[p^2+ z^2]^\alpha} \right] =&\Re \left[ i \int_0^1 \mathrm{d}p p^{d-1} \int_0^p \frac{\mathrm{d}z}{[p^2- z^2]^\alpha} \right] \\
&+\Re \left[ i \int_0^1 \mathrm{d}p p^{d-1} \int_p^1 \frac{\mathrm{d}z}{[p^2- z^2]^\alpha} \right].
\end{split}
\end{equation}
The first row of Eq.~\eqref{eq:lineintegralhierarchies} vanishes (with convergent parameter values), as the integral is purely imaginary, evaluating to
\begin{equation}
i \int_0^1 \mathrm{d}p p^{d-1} \int_0^p \frac{\mathrm{d}z}{[p^2- z^2]^\alpha} = \frac{i}{2} \left[\int_0^1 \mathrm{d}p p^{d-2 \alpha}  \right] \left[\int_0^1 \frac{\mathrm{d}w}{\sqrt{w} (1-w)^\alpha} \right]\\
= \frac{i}{2(d-2\alpha+1)} \frac{\Gamma \left(\frac{1}{2} \right)\Gamma \left(1-\alpha\right)}{\Gamma \left(\frac{3}{2}-\alpha \right)}.
\end{equation}
The second row of Eq.~\eqref{eq:lineintegralhierarchies} is generally non-vanishing, as we can observe that it contains an overall $e^{i \pi \alpha}$. Explicitly one finds 
\begin{equation}
\begin{split}
 i\int_0^1 \mathrm{d}p p^{d-1} \int_p^{1} \frac{\mathrm{d}z}{[p^2-z^2]^\alpha}= \frac{i\pi [i + \cot (\alpha \pi) ] \Gamma \left(\alpha-\frac{d}{2} \right)}{d(d-2\alpha+1) \Gamma (\alpha) \Gamma \left(-\frac{d}{2} \right)} \frac{\sin \left(\pi\alpha -\frac{\pi d}{2} \right)}{\sin \left(\frac{\pi d}{2} \right)},
\end{split}
\end{equation}
which we notice to converge numerically (even) when $\alpha < 1/2$ for the full  range of dimensional values allowing convergence at all. More specifically, one can recognize that the special functions used require $0< \frac{d}{2} < 1$ and $d+1 > 2 \alpha > 0$, which in turn can be analytically continued to almost the full real axis. For the real part, we get, after some manipulations, 
\begin{equation}
 I_{\Sigma, \text{left}} = \Re \left[ i\int_0^1 \mathrm{d}p p^{d-1} \int_p^{1} \frac{\mathrm{d}z}{[p^2-z^2]^\alpha} \right]= \frac{\pi \Gamma \left(\frac{d}{2} \right)}{2(d-2\alpha+1)\Gamma(\alpha)\Gamma \left(\frac{d}{2}-\alpha+1 \right) }.
 \label{eq:vert_int}
\end{equation}
The expression for the topmost line integral in $\Sigma$ is then found from Eq.~\eqref{eq:sigma_vanish} as $-I_{\Sigma, \text{lower}}-I_{\Sigma, \text{left}}$. 
This expression is found to be vanishing at $\alpha \rightarrow 1$, which indeed agrees with the fact that $I_1^t(\mu)$ reproduces $I_1(\mu)$. 
This agreement reassures us that we can treat the neighborhood of $\alpha = 1$ in a consistent manner, and yields insight into the  piecewise behavior of $I_\alpha$ when $\alpha$ tends towards an integer value. 

The computation is finished by combining contributions arising from both $\theta(\mu-p)$ and $\theta(p-\mu)$, i.e., the residue result and the supplementary correction. Since the $-I_{\Sigma, \text{lower}}$ piece from the $p<\mu$ contribution cancels the $p>\mu$ contribution completely, the full result is just the vertical, $-I_{\Sigma, \text{left}}$ piece from the $p<\mu$ contribution, namely
\begin{equation}
I_\alpha(\mu, T) \underset{T \rightarrow 0}{\longrightarrow} 
-\left(  \frac{e^{\gamma_\text{E}} \Lambda^2}{4 \pi}\right)^\epsilon\frac{ 1  }{ (4 \pi)^\frac{d}{2} \Gamma (\alpha) \Gamma \left(\frac{d}{2}+1-\alpha \right)}\frac{\mu^{d+1-2 \alpha}}{(d+1-2 \alpha)},
\end{equation}
in full agreement with the result from Eq.~\eqref{eq:zero-result}. This completes our demonstration that $I_\alpha(\mu)$ can be correctly (with respect to physics) evaluated for all $\alpha$ with either order of integration.

\section{Results}
\label{sec:results}
\noindent
In this work, we have explored an apparent ambiguity that takes place in loop integrals with finite chemical potential (imaginary scale) at zero temperature. As seen through the simplest relevant example, by integrating first over $p_0$ and applying the residue theorem, one obtains results that disagree with those obtained by first integrating over the dimensionally regularized spatial integral. The latter can also be seen as the proper extension from finite temperature,  while the former is algebraically much more attractive for most computations. As such, it is well-motivated to seek ways to augment the results from residue theorem, and apply these lessons to more challenging integral structures.

By considering integer-valued exponents in the loop integral of interest, we find the difference between the two methods appearing as a delta-function contribution at the edge of the cut-off ($\Im(p^0) = \mu$) arising from the residue theorem. This implies that the boundary at $\Im(p^0) = \mu$ in some sense experiences differentiation, akin to the other elements in the integrand (contrary to what naive use of the residue theorem suggests). This can be understood to arise from the Fermi--Dirac distribution functions in finite temperature expressions, or alternatively from the fundamental symmetries of dimensional regularization as an iterative process (IBP). The significance of the boundary $\Im(p^0) = \mu$ is even more prominent when considering non-integer exponents. In that case, we see the line integral along this edge fully generates the novel structure, seen to arise from dimensional regularization. 

Moreover, we find a major difference in what regions the integral discards, depending on which way we integrate. 
Specifically, evaluating the spatial integral first yields an expression covering all of the $p \times p_0$ space  $\mathbb{R}_+ \times \mathbb{R}$. By contrast, generalizing the residue result in a naive fashion rejects the region $[0,\mu]\times \mathbb{R}$ from the full integration region, as described above (which then receives the boundary correction from the previous paragraph, to obtain the physically motivated  answer). However, this is exclusive to integer-valued exponents; in the case of non-integer exponents there are contributions from the region $[0,\mu]\times \mathbb{R}$, which do not appear from our residue-driven prescription. 

We have demonstrated that the Feynman integrals with first-order poles behave in the same manner in either order of integration, in agreement with \cite{ghisou}. We have also demonstrated that higher-order  results can be related to the zero-temperature limits of finite-temperature expressions supplemented by iterative differentiation of propagators (being careful not to move any regulatory limits outside the outermost $p$ integral), which is highly beneficial for more versatile multi-loop integrals. Explicitly, 
\begin{equation}
\int_p \frac{1}{(\widetilde{P}^2)^k} \equiv \int_p \frac{1}{[(p_0+i\mu)^2+p^2]^{k}} = \left[\int_p\underset{E_p \rightarrow p^+}{\text{lim}} (-1)^{k-1} \frac{\mathrm{d}^{k-1}}{\mathrm{d}\!\left(E^2_p\right)^{k-1}} \right]  \frac{1}{[(p_0+i\mu)^2+E_p^2]},
\end{equation}
in which the tildes denote a shift in the temporal component by $i \mu$. Such a prescriptions aids even in a computation as simple as
\begin{equation}
\begin{split}
    \int_P \frac{1}{\tilde{P}^4 (\tilde{P}+K)^2} &\equiv \int_p \int_{-\infty}^\infty \frac{\mathrm{d}p_0}{2 \pi}\frac{1}{[(p_0+i\mu)^2+p^2]^2 [(p_0+k_0+i\mu)^2+|\mathbf{p}+\mathbf{k}|^2]}\\
    &\mapsto -\int_p \underset{E_p \rightarrow p^+}{\text{lim}}\left\{\frac{\mathrm{d}}{\mathrm{d}\! \left( E_p^2 \right)} \int_{-\infty}^\infty \frac{\mathrm{d}p_0}{2 \pi} \frac{1}{[(p_0+i\mu)^2+E_p^2]  [(p_0+k_0+i\mu)^2+|\mathbf{p}+\mathbf{k}|^2]} \right\}.
    \end{split}
\end{equation}
Here the added \emph{thermal} contribution (neglected by simply using the residue theorem) would be seen through $\frac{1}{2 E_p} \frac{\mathrm{d}}{\mathrm{d}E_p} \theta(E_p-\mu)$. Furthermore, we wish to emphasize here the importance of the choice of differentiation variable, as some alternate formulations with e.g. $p^2 \mapsto p^2 + m^2$ and acting with $\frac{\mathrm{d}}{\mathrm{d} (m^2)}$ might indicate that differentiation could take place outside the spatial integral. Particularly in multi-loop computations, this can lead to situations where the added scale $m^2$ regulates the expression in places where dimensional regularization should have done so. Upon differentiation this can lead to expressions that are nearly impossible to relate to the correct result. 

Additionally, we want to emphasize that the procedure is only safe with expressions which formally allow the application of the residue theorem. This can for the most part be remedied by moving divergent elements to the spatial region of the integration. Consider for example
\begin{eqnarray}
\int_P \frac{\tilde{p}_0^4}{\tilde{P}^4} = \int_P 1 + \int_P \frac{p^4}{\tilde{P}^4}-2 \int_P   \frac{p^2}{\tilde{P}^2} \mapsto \int_P \frac{p^4}{\tilde{P}^4}-2 \int_P   \frac{p^2}{\tilde{P}^2},
\end{eqnarray}
where we re-cast the expression such that the (augmented) residue theorem can be applied to the $p_0$ integral with ease. In the final decomposition, we have removed all $p_0$ terms from the numerators (as they would modify the analytic structure), and the remaining integrals are regulated by the $d$ dimensional spatial integral.

\section{Discussion}
\label{sec:discussion}
\noindent
We have described some properties of complex-valued dimensionally regulated integrals encountered in quantum field theory, observing that changing the order of integration is only possible by augmenting the residue theorem with boundary terms. While our treatment has been largely formal, the relevant integrals are present in physical systems. An example of this is in high-density zero-temperature quantum chromodynamics: At high orders in perturbation theory, performing the temporal integrals associated with zero-components of the momenta as a first step becomes increasingly appealing. 

We have observed that doing so by only taking residues misses, at least in the general case, certain physically motivated contributions when propagators with an exponent $\alpha>1$ are present---that is to say, an application of the residue theorem in finite-density computations is only possible for integrals that converge properly. Even the simple one-loop integral we have carefully studied displaying this problematic behavior is required for \emph{general real} exponents starting at next-to-next-to-leading order in perturbative quantum chromodynamics, and integrals with nontrivial exponents are increasingly commonplace at higher orders. 

Performing spatial integrals first alleviates the problem by immediately introducing a suitable regulator, and the problem was also seen to be absent at finite temperatures, with the problem arising with a non-careful treatment of the zero-temperature limit. However, neither approach may be feasible in all situations, and as such alternative approaches are called for. 

A few possibilities to properly include all contributions have been discussed above. They include adding in the missing (boundary) terms, inspired by careful contour integration and arising from derivatives of step functions; and decreasing the exponent of an integer-valued propagator by taking derivatives. In the latter case one must be careful with the commutativity of limits: a derivative with respect to an auxiliary mass parameter appearing in a propagator might not commute especially with the spatial integrals, and failing to take this into account can run the risk of introducing new divergences associated with the auxiliary mass parameter. To summarize, we emphasize the need for caution in computations involving the use of the residue theorem via an interchange of integration orders when evaluating divergent expressions, as the theorem may lead to incorrect physics without additional boundary terms.

\section*{Acknowledgments}
\noindent
The authors would like to thank Aleksi Vuorinen for enlightening discussions and feedback on earlier versions of this manuscript. We would also like to thank the anonymous referee for their insightful and extensive feedback. TG was supported in part by the Deutsche Forschungsgemeinschaft (DFG, German Research Foundation) -- Project-ID 279384907 -- SFB 1245 and by the State of Hesse within the Research Cluster ELEMENTS (Project ID 500/10.006).
JÖ acknowledges financial support from the Vilho, Yrjö and Kalle Väisälä Foundation of the Finnish Academy of Science and Letters. 

\bibliographystyle{unsrt}
\bibliography{referenceloop.bib}

\end{document}